\documentclass[twocolumn, pra,showpacs,superscriptaddress]{revtex4}
\usepackage{graphicx}
\usepackage{dcolumn}
\usepackage{bm}
\usepackage{amsmath}

\begin{document}

\title{Dynamical properties of hard-core anyons in one-dimensional optical lattices}
\author{Yajiang Hao}
\email{haoyj@ustb.edu.cn}
\affiliation{Department of Physics,
University of Science and Technology Beijing, Beijing 100083, China}
\author{Shu Chen}
\email{schen@aphy.iphy.ac.cn}
\affiliation{Beijing National
Laboratory for Condensed Matter Physics, Institute of Physics,
Chinese Academy of Sciences, Beijing 100080, China}
\date{\today}

\begin{abstract}
We investigate the dynamical properties of anyons confined in
one-dimensional optical lattice combined with a weak harmonic trap
using the exact numerical method based on a generalized
Jordan-Wigner transformation. The density profiles, momentum
distribution, occupation distribution and occupations of the
lowest natural orbital are obtained for different statistical
parameters. The density profiles of anyons  display the same
behaviors irrespective of statistical parameter in the full
evolving period. While the behaviors dependent on statistical
property are shown in the momentum distributions and occupations
of natural orbitals.
\end{abstract}
\maketitle
% 05.30.Pr Fractional statistics systems
% 03.75.Kk Dynamic properties of condensates; collective and hydrodynamic excitations,  superfluid flow
% 05.30.Fk Fermion systems and electron gas
% 05.30.Jp Boson systems
% 67.10.Fj Quantum statistical theory

%\preprint{APS/123-QED}

\narrowtext

\section{introduction}

Generally according to the quantum statistical property particle
is classified as boson and fermion. The  wavefunction of identical
bosons is symmetric under exchange while that of identical
fermions is antisymmetric under exchange. As a natural
generalization, physicist proposed that there exists anyon
interpolating between Bose and Fermi statistics, which satisfies
fractional statistics \cite{anyons}. It has become an important
concept in condensed matter physics
\cite{Laughlin,Halperin,Camino} and has ever been used for
successfully explaining the fractional quantum Hall effect (FQHE)
\cite{SuperConductor}. Fractional statistics also play important
roles in the theory for non-fermi liquid, Chern-Simon theory, and
other aspects \cite{YSWu,ZNCHa}. Another potential application of
anyons is to realize the topological quantum computation with
non-abelian anyons \cite{RMP2008,Kitaev}. Besides the traditional
two dimensional electron system, cold atoms in low dimension is
also a popular platform to realize the fractional statistics. It
has been suggested that anyons can be created, detected and
manipulated in rotating Bose-Einstein condensates (BECs) and cold
atoms in optical lattices \cite{PZoller,DuanLM,OL}. Particularly,
researches on anyons in cold atoms are not restricted in two
dimensional system but the proposal to realize anyons in one
dimensional (1D) optical lattice is also put forward recently
\cite{NatureComm}.

Since BECs are realized experimentally, cold atoms has been making
great progress both experimentally and theoretically for its
``purity" and high controllability comparing with the traditional
condensed matter system. For the dimensionality effect and
extremely profound strong correlation in low dimensional system,
the quantum gas in low dimension has also attract much attention
\cite{1D}. With anisotropic magnetic trap or two-dimensional
optical lattice potentials, the particle motion is tightly
confined in two directions to zero point oscillations
\cite{Ketterler,Paredes,Toshiya} and the strongly correlated
Tonks-Girardeau (TG) regime can be achieved
\cite{Paredes,Toshiya}. By crossing the confinement-induced
resonance (CIR) from the TG gas, the super TG (sTG) gas is
accessible, which show stronger correlation than TG gas
\cite{STG}. The interaction between atoms can be tuned in the full
interacting regime with Feshbach resonance technique and
confinement-induced resonance by tuning magnetic field. It is the
excellent tunability of cold atom that makes it to be an excellent
candidate to realize fractional statistics.

In fact, before the experimental scheme of  Keilmann et. al.
\cite{NatureComm}, 1D anyon gas has been investigated
theoretically in various 1D systems \cite
{Haldane,WangZD,Kundu99,Girardeau06} including the Bose quantum
gas. Kundu proved that a 1D Bose gas interacting through
$\delta$-function potential combined with double $\delta$-function
potential and derivative $\delta$-function potential is equivalent
to the anyon gas interacting via $\delta$-function potential
\cite{Kundu99}. This stimulated many research interests on
$\delta$-anyon gas
\cite{Girardeau06,Batchelor,Patu07,Calabrese,anyonTG,Patu08,Cabra,Campo,HaoPRA78,HaoHCA79,Amico}.
It turns out that the ground state density distribution of
$\delta$-anyon gas displays similar behavior as that of Bose gas
with the increasing interaction. In the strong interacting regime
the density distribution shows the same behavior as the free
fermion \cite{Girardeau,Hao06,Hao07,Zoellner,Deuretzbacher}, which
is irrespective to the statistical parameter. The special property
resulted from the fractional statistics exhibits in the reduced
one body density matrices and the momentum distributions
\cite{Lenard,Vaidya}. The momentum distribution of anyon differs
from fermion's oscillations and boson's single peak structure,
which are symmetric about the zero momentum. The momentum
distribution of anyons is asymmetric when the statistical
parameters deviate from the Bose and Fermi limit
\cite{HaoPRA78,HaoHCA79,anyonTG,Patu08,Campo}. This special
behavior originates from that the reduced one body density matrix
is a complex Hermitian one rather than a real one for Boson and
Fermion.

While most studies have focused on the static properties of the 1D
anyonic gas, its dynamics remains to be investigated. The present
paper shall study the dynamics of anyons confined in optical
lattice with weak harmonic trap in the hard core limit. In Ref.
\cite{HaoHCA79}, we have extended the exact numerical method
originally used to treat hard-core bosons by Rigol {\it et al.}
\cite{MRigol} to deal with the ground state of hard-core anyons
(HCA) in optical lattice. Here, we further extend this method to
investigate the dynamics of HCA. By evaluating the exact
time-dependent one-particle Green's function, we obtain the
reduced one body density matrix (ROBDM) and thus the density
profiles and the momentum distribution at arbitrary time. The
dynamical properties induced by anyonic statistics shall be
displayed in the momentum distribution.

The paper is organized as follows. In Sec. II, we give a brief
review of 1D anyonic model and introduce the numerical method. In
Sec. III, we present the density profiles, momentum distributions, occupation distribution and
the occupations of the lowest natural orbital for different statistics parameters.
A brief summary is given in Sec. IV.

\section{formulation of the model and method}

We consider $N$ hard core anyons confined in an optical lattice of
$L$ sites with a weak harmonic trap and the second quantized
Hamiltonian can be formulated as
\begin{eqnarray}
H=-t\sum_{l=1}^L\left( a_{l+1}^{\dagger }a_l+H.C.\right)
+\sum_{l=1}^LV_la_l^{\dagger }a_l.
\end{eqnarray}
Here the harmonic potential $V_l=V_0(l-(L+1)/2)^2$ with the strength
of the harmonic trap $V_0$. The anyonic operator $a_l^{\dagger }$
($a_l$) create (annihilate) an anyon on site $l$ and satisfies the
generalized commutation relations
\begin{eqnarray}
a_ja_l^{\dagger } &=&\delta _{jl}-e^{-i\chi \pi \epsilon
(j-l)}a_l^{\dagger }a_j, \nonumber\\
a_ja_l &=&-e^{i\chi \pi \epsilon (j-l)}a_la_j
\end{eqnarray}
for $j\neq l$, where the sign function $\epsilon (x)$ gives -1, 0
or 1 depending on whether $x$ is negative, zero, or positive and
$\chi$ is the parameter related with fractional statistics. The
generalized commutation relations reduce to fermionic commutation
for $\chi =0$ and reduce to Bose commutation for $\chi =1$, while
for anyons satisfying fractional statistics $\chi$ changes in
between them. The hard core interactions between anyons restricts
the additional condition $a_l^2=a_l^{\dagger 2}=0$ and $\left\{
a_l,a_l^{\dagger }\right\} =1$. In the Hamiltonian $t$ denotes the
hopping between the nearest neighbour sites, which can be tuned by
changing the strength of optical lattice.

In order to solve the model of hard core anyons, we extend the
numerical method to investigate the hard core bosons in optical
lattice developed by Rigol \emph{et. al.}. We can map the above
model into the polarized fermionic Hamiltonian using the
generalized Jordan-Wigner transformation
\begin{eqnarray}
a_j &=&\exp \left( i\chi \pi \sum_{1\leq s<j}f_s^{\dagger }f_s\right) f_j, \\
a_j^{\dagger } &=&f_j^{\dagger }\exp \left( -i\chi \pi \sum_{1\leq
s<j}f_s^{\dagger }f_s\right),
\end{eqnarray}
where $f_j^{\dagger }$ ($f_j$) is creation (annihilation) operator
for fermions. The above Hamiltonian of system with $N$ anyons is
transformed into a fermionic one with $N_F=N$ fermions
\begin{eqnarray}
H_F =-t\sum_{l=1}^L\left( f_{l+1}^{\dagger }f_l+H.C.\right)
+\sum_{l=1}^LV_lf_l^{\dagger }f_l,
\end{eqnarray}
where the Fermionic operator satisfy the Fermi anti-commutation
relation
\begin{eqnarray}
\left\{ f_i,f_j^{\dagger }\right\} =\delta _{ij},\left\{
f_i,f_j\right\} =\left\{ f_i^{\dagger },f_j^{\dagger }\right\} =0.
\end{eqnarray}
The original question about anyons now can be investigated by
solving the model on the polarized fermions in optical lattice. We
can obtain the exact many body wavefunction of polarized fermions
with the diagonalized method and therefore the ground state and
interesting physical phenomena of hard core anyons.

The equal-time Green's function for the hard core anyons at time
$\tau$ should be expressed as
\begin{eqnarray}
G_{jl}\left( \tau \right) &=&\left\langle \Psi _{HCA}\left( \tau
\right) \left| a_ja_l^{\dagger }\right| \Psi _{HCA}\left( \tau
\right) \right\rangle
\\  \nonumber
&=&\left\langle \Psi _F\left( \tau \right) \right| \exp \left( i\chi \pi
\sum_\beta ^{j-1}f_\beta ^{\dagger }f_\beta \right) f_jf_l^{\dagger } \\    \nonumber
&&\times \exp \left( -i\pi \sum_\gamma ^{l-1}f_\gamma ^{\dagger
}f_\gamma
\right) \left| \Psi _F\left( \tau \right) \right\rangle \\  \nonumber
&=&\left\langle \Psi _F^A\left( \tau \right) |\Psi _F^B\left( \tau \right)
\right\rangle
\end{eqnarray}
with
\begin{eqnarray*}
\left\langle \Psi _F^A\left( \tau \right) \right| &=&\left(
f_j^{\dagger }\exp \left( -i\chi \pi \sum_\beta ^{j-1}f_\beta
^{\dagger }f_\beta \right)
\left| \Psi _F\left( \tau \right) \right\rangle \right) ^{\dagger }, \\
\left| \Psi _F^B\left( \tau \right) \right\rangle &=&f_l^{\dagger
}\exp \left( -i\chi \pi \sum_\gamma ^{l-1}f_\gamma ^{\dagger
}f_\gamma \right) \left| \Psi _F\left( \tau \right) \right\rangle .
\end{eqnarray*}
Here $\left| \Psi _{HCA}\left( \tau \right) \right\rangle$ is the
wavefunction at time $\tau$ of hard core anyons in a system with
Hamiltonian $H_{HCA}$, and $\left| \Psi _F( \tau ) \right\rangle$ is
the corresponding one for the equivalent polarized fermions. For the
polarized fermions the time evolution of their initial wavefunction
$\left| \Psi _F^I\right\rangle$
\begin{eqnarray}
\left| \Psi _F\left( \tau \right) \right\rangle =e^{-i\tau H_F/\hbar
}\left| \Psi _F^I\right\rangle.
\end{eqnarray}
While the matrix representation of initial wavefunction can be
expressed as
\begin{eqnarray}
\left| \Psi _F^I\right\rangle
=\prod_{n=1}^{N_f}\sum_{l=1}^LP_{ln}^If_l^{\dagger }\left|
0\right\rangle
\end{eqnarray}
so that
\begin{eqnarray*}
\left| \Psi _F\left( \tau \right) \right\rangle &=&e^{-i\tau
H_F/\hbar }\prod_{n=1}^{N_f}\sum_{l=1}^LP_{ln}^If_l^{\dagger
}\left| 0\right\rangle \\
&=&\prod_{n=1}^{N_f}\sum_{l=1}^LP_{ln}\left( \tau \right)
f_l^{\dagger }\left| 0\right\rangle.
\end{eqnarray*}
Thus the fermonic time-dependent wavefunction can be expressed as an
$L\times N_f$ matrix ${\bf P}(\tau)$. The matrix ${\bf P}\left( \tau
\right) $ can be evaluated as
\begin{eqnarray}
e^{-i\tau H_F/\hbar }{\bf P}^I=Ue^{-i\tau D/\hbar }U^{\dagger }{\bf
P}^I,
\end{eqnarray}
where $U$ is an unitary transformation diagonalizing the Hamiltonian
$H_F$, \emph{i.e.}, $U^{\dagger }H_FU=D$ with diagonal matrix $D$.
After an easy evaluation the state $\left| \Psi _F^A\right\rangle $
reads
\[
\left| \Psi _F^A\left( \tau \right) \right\rangle
=\prod_{n=1}^{N_f+1}\sum_{l=1}^LP_{ln}^{\prime A}\left( \tau \right)
f_l^{\dagger }\left| 0\right\rangle
\]
with
\begin{eqnarray*}
P_{ln}^{\prime A}\left( \tau \right) &=&\exp \left( -i\chi \pi \right)
P_{ln}\left( \tau \right)
\begin{array}{lll}
&  &
\end{array}
\text{for }l\leq j-1 \\
P_{ln}^{\prime A}\left( \tau \right) &=&P_{ln}\left( \tau \right)
\begin{array}{llllllllllll}
&  &  &  &  &  &  &  &  &  &  &
\end{array}
\text{for }l\geq j
\end{eqnarray*}
for $n\leq N_f$ and $P_{jN_f+1}^{\prime A}\left( \tau \right) =1$
and $ P_{lN_f+1}^{\prime A}\left( \tau \right) =0$ $\left( l\neq
j\right) $. The state $\left| \Psi _F^B\right\rangle $ has the same
form with the replace of $j$ by $l$. The time-dependent Green's
function is a determinant dependent on the $L\times \left(
N_f+1\right) $ matrixes ${\bf P}^{\prime A}\left( \tau \right) $ and
$ {\bf P}^{\prime B}\left( \tau \right) $
\begin{eqnarray*}
G_{jl}\left( \tau \right) &=&\left\langle \Psi _F^A\left( \tau
\right) |\Psi _F^B\left( \tau \right) \right\rangle =\det \left[
\left( {\bf P}^{\prime A}\left( \tau \right) \right) ^T{\bf P}
^{\prime B}\left( \tau \right) \right] .
\end{eqnarray*}
In the present paper we will focus on the time evolution of the
density profile and momentum distribution for hard core anyons with
different statistical parameter $\chi$. The ROBDM can be evaluated
by Green's function
\begin{eqnarray*}
\rho _{jl}\left( \tau \right) =\left\langle a_j^{\dagger
}a_l\right\rangle =\delta _{jl}\left( 1-G_{jl}\left( \tau \right)
\right) -(1-\delta _{jl})e^{-i\chi \pi }G_{jl}\left( \tau \right).
\end{eqnarray*}
The diagonal part of ROBDM is the density profile and its Fourier
transformation is defined as momentum distribution
\begin{eqnarray}
n(k)=\frac{1}{2\pi}\sum_{j,l=1}^Le^{-ik(j-l)}\rho _{jl}(\tau).
\end{eqnarray}
The natural orbitals $\phi ^{\eta}$ are defined as the
eigenfunctions of the one-particle density matrix
\begin{eqnarray}
\sum_{j=1}^L\rho _{jl}\phi ^{\eta}=\lambda _{\eta}\phi ^{\eta},
j=1,2,...L,
\end{eqnarray}
and can be understood as the effective single-particle states with
occupations $\lambda _{\eta}$.

\section{dynamics of density profile and momentum distribution}

In the present paper we investigate the dynamics of hard core
anyons in optical lattice. Initially we confine the anyons in
optical lattice with a strong harmonic trap and then turn off the
harmonic trap or reduce the strength of harmonic trap. The trapped
anyons will evolve in the optical lattice. For convenience we set
lattice constant $a$ as 1, the unit of $k$ is $1/a$ and the unit
of time is $\hbar/t$.

\begin{figure}[tbp]
\includegraphics[width=3.5in]{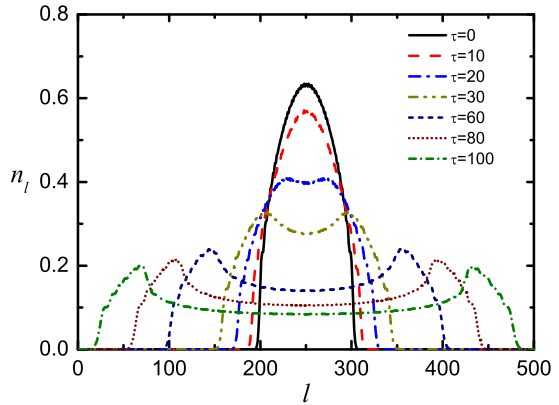}
\caption{(color online) The evolving of density distribution for 50
hard core anyons in optical lattice of 500 sites.
$V_0^I=1.0\times10^{-3}t$ and $V_0=1.0\times10^{-8}t$.} \label{fig1}
\end{figure}
In Fig. 1, 2, 3 and 4 we show the evolution of 50 anyons in
optical lattice of 500 sites. Initially the strength of harmonic
trap $V_0^I=1.0\times 10^{-3}t$ and at $\tau=0$ the harmonic trap
is released to $V_0(0)=1.0\times 10^{-8}t$, i.e., the anyons are
trapped only in optical lattice. The evolution of density distribution is shown in Fig. 1. It turns out that the density
distribution of anyons shall not display any different dynamical
property from those of bosons and fermions. We cannot distinguish
the statistical properties by the density distribution in real
space. Similar to hard core bosons and polarized fermions, anyons
expand in the optical lattice and they shall gradually populate in
the full lattice. The density of anyons at the center reduces
faster than the density at the border because the anyons locating
in the sites of high density posses higher energy. As evolving
time is long enough anyons homogeneously distribute in the middle
regime while its density distribution shows peaks at the border
regime.

\begin{figure}[tbp]
\includegraphics[width=3.5in]{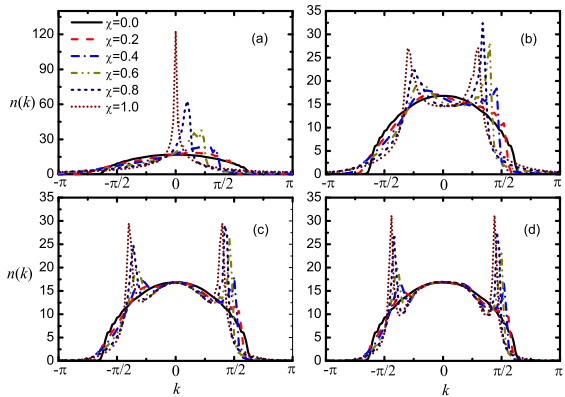}
\caption{(color online) The momentum distributions for 50 hard core anyons
in lattice of 500 sites, $V_0^I=1.0\times10^{-3}t$ and
$V_0=1.0\times10^{-8}t$. (a) $\tau=0$, (b) $\tau=20$, (c)
$\tau=50$,(d) $\tau=100$.
%For convenience we set lattice constant $a=1$ and the unit of $k$ is $1/a$. Times ($\tau$) are in units of
%$\hbar/t$.
} \label{fig2}
\end{figure}
The corresponding momentum distribution is displayed in Fig. 2.
Initially bosons and fermions distribute symmetrically about the
zero momentum as respective statistical property and anyons
($0<\chi<1$) exhibit the asymmetrical momentum distribution. After
the harmonic trap is turned off for anyons of different
statistical parameter their momentum distribution show different
evolving properties. Fermions ($\chi = 0.0$) do not show obvious
change of momentum distribution (In fact the slight change happens
according to the numerical data). As the statistical parameter
deviates from $\chi=0$ the momentum distribution of anyon evolves
from the asymmetrical structure of single peak to the structure
similar to that of Fermions but with two asymmetrical peaks. As
statistical parameter increases these two peaks become more and
more obvious. For the case of $\chi =1$ (hard core bosons) the
momentum distribution evolve to the structure of symmetrical
double peaks. When the evolving time is long enough the momentum
distribution of anyons of any statistical parameter ($0\leq \chi
\leq 1$) exhibits identical behavior in the middle regime.

\begin{figure}[tbp]
\includegraphics[width=3.5in]{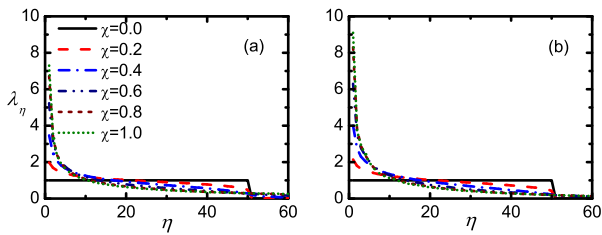}
\caption{(color online) The occupation distributions for 50 hard core anyons
in lattice of 500 sites, $V_0^I=1.0\times10^{-3}t$ and
$V_0=1.0\times10^{-8}t$. (a) $\tau=0$, (b) $\tau=100$.
%For convenience we set lattice constant $a=1$ and the unit of $k$ is
%$1/a$. Times ($\tau$) are in units of $\hbar/t$.
} \label{fig3}
\end{figure}
\begin{figure}[tbp]
\includegraphics[width=3.0in]{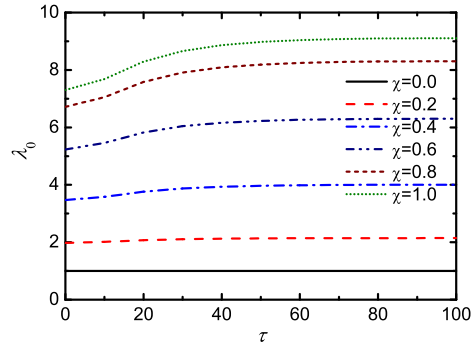}
\caption{(color online) The occupations of the lowest natural
orbital for 50 bosons in lattice of 500 sites.
$V_0^I=1.0\times10^{-3}t$ and $V_0=1.0\times10^{-8}t$.} \label{fig4}
\end{figure}
In Fig. 3 we show the occupation distribution of natural orbitals
for the same system as above. In the full evolution period the
occupation distribution of each orbital does not change
qualitatively. In the Bose limit anyons occupy the lower natural
orbitals and the occupation distribution displays the single peak
structure. In the Fermi limit each anyon occupies one natural
orbital and at any time the occupation distribution seems like a
step-function. While for anyons in between these two limits, the
occupations of higher natural orbitals increase as the statistical
parameter evolves from Bose limit to Fermi limit. We also display
the evolution of the occupation of the lowest natural orbital in
Fig. \ref{fig4}. It is shown that the occupation is time
independent in the Fermi limit and as deviating from the Fermi
limit the occupation increases during the time evolution. The
increase is more obvious for the bigger statistical parameter.
When the evolving time is long enough the occupation of the lowest
natural orbital shall preserve a constant.

\begin{figure}[tbp]
\includegraphics[width=3.5in]{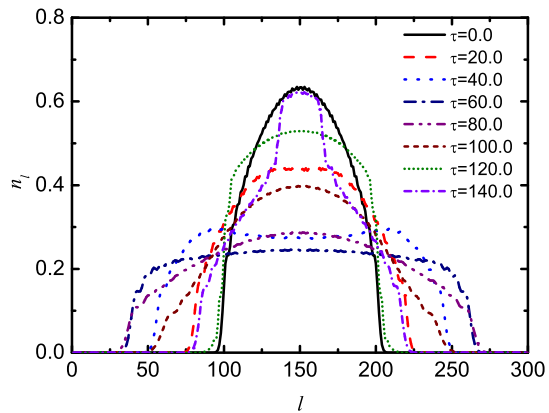}
\caption{(color online) The density distribution of 50 hard core
anyons in optical lattice of 300 sites. $V_0^I=1.0\times10^{-3}t$
and $V_0=2.0\times10^{-4}t$.} \label{fig5}
\end{figure}
If the initially confined hard core anyons in optical lattice
evolve in a weak harmonic trap rather than turning off the
harmonic potential, the situation will be different. Anyons with
different statistical parameters always exhibit the same density
distributions and we also cannot determine the statistical
properties according to evolving property of density profiles. In
Fig. \ref{fig5} we display the evolving density distribution of 50
anyons in optical lattice of 300 sites combined with a weaker
harmonic potential ($V_0=2.0\times10^{-4}t$). Initially, anyons
distribute in the middle regime of the harmonic trap and after the
harmonic potential becomes weak anyons shall expand firstly. The
central density decreases faster than the boundary density such
that the density profiles behave as a Fermi-like distribution at
$\tau=60$. Then anyons shall contract because of the confinement
of harmonic trap. With the time evolution anyons redistribute in
the middle regime of the trap.

\begin{figure}[tbp]
\includegraphics[width=4.0in]{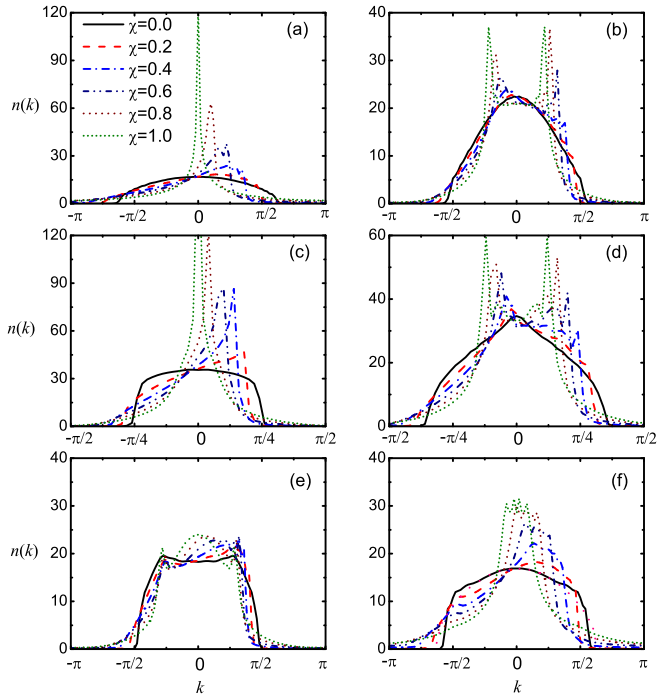}
\caption{(color online) The momentum distribution of 50 anyons in
optical lattice of 300 sites. $V_0^I=1.0\times10^{-3}t$ and
$V_0=2.0\times10^{-4}t$. (a) $\tau=0$, (b) $\tau=30$, (c) $\tau=60$,
(d) $\tau=80$, (e) $\tau=100$, (f) $\tau=120$.} \label{fig6}
\end{figure}
During the expansion, momentum distributions show rich dynamical
structure (Fig. \ref{fig6}). In the Bose limit, the momentum
distribution firstly evolves from the original structure of a
single peak to the structure of double peaks, and then back to the
structure of a single peak at $\tau=60$. During the later period
momentum distribution shall behave as the single peak and double
peaks alternately. In the Fermi limit the momentum distribution
does not keep its original profile as shown in Fig. \ref{fig2}  and shall exhibit the oscillating behavior.
Contrary to the evolution of density distribution,
%the expansion first and then contraction of density
%distribution,
the momentum distribution contracts firstly and at $\tau=60$ (Fig.
\ref{fig6}c) displays the step-function profile in the region of
$-\pi/4 \leq k\leq \pi/4$. Then it gradually expands to the region
of higher momentum and at $\tau=120$ (Fig. \ref{fig6}f) momentum
distribution almost recovers to the behavior at $\tau=0.0$. For
anyon gas in between ($0<\chi<1$) its asymmetric momentum
distribution also shows the alternate behavior. For the case of
bigger statistical parameter asymmetric double peaks are displayed
and for the case of smaller statistical parameter momentum
distribution exhibits the behavior similar to those of fermions.

\begin{figure}[tbp]
\includegraphics[width=3.5in]{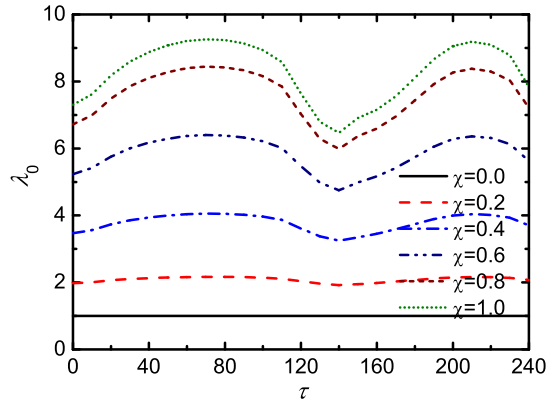}
\caption{(color online) The occupation of the lowest natural orbital
for 50 anyons in optical lattice of 300 sites.
$V_0^I=1.0\times10^{-3}t$ and $V_0=2.0\times10^{-4}t$.} \label{fig7}
\end{figure}
The evolution of occupation for the anyons in optical lattice with
weak harmonic potential is similar to the case only confined in
optical lattice. The occupation distribution always exhibits the
same behaviors qualitatively as those at initial time. But the
occupation of the lowest natural evolve with time, which is
displayed in Fig. \ref{fig7}. It is shown that in the Fermi limit
the occupation is time-independent and as deviating from the Fermi
limit it shall oscillate with the time evolution. It does not
increase monotonously rather than fluctuates with the time
evolution. The bigger the statistical parameter, the stronger the
oscillation amplitude.

\begin{figure}[tbp]
\includegraphics[width=3.5in]{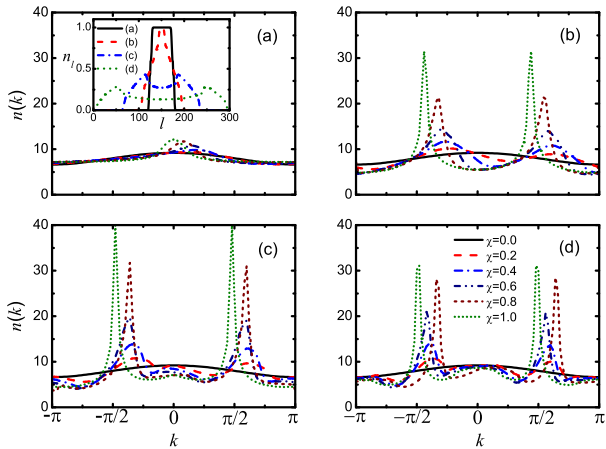}
\caption{(color online)  The momentum distribution of 50 anyons in
lattice of 300 sites. $V_0^I=1.0\times10^{-2}t$ and
$V_0=1.0\times10^{-8}t$. (a) $\tau=0$, (b) $\tau=10$, (c)
$\tau=30$,(d) $\tau=60$. Inset: Density distributions for the same
system.} \label{Mott}
\end{figure}
In order to investigate the dynamical properties of anyons in Mott
regime, we initially prepare a Mott state by superimposing optical
lattice with a tight harmonic potential and then turn off it at
$\tau=0.0$. The dynamical evolutions of Mott state are displayed
in Fig. \ref{Mott}. Initially, the momentum distribution for
anyons of any statistical parameter exhibits the behavior similar
to that of fermions. There are only tiny differences in the regime
close to zero momentum. After the harmonic trap is turned off, the
momentum distribution of anyons in the Fermi limit always preserve
its initial profile. Anyons deviating this limit still display
Fermi-like momentum distribution in the full momentum regime, but
the structure of double peaks appears at particular regime. In the
full evolving period anyons in Bose limit show the sharpest peaks
and as statistical parameter decreases
 (close to the Fermi limit) the peaks become smaller. It is
same as before that anyons always exhibit the asymmetrical
momentum distributions except of the Bose limit and Fermi limit.
We also display the evolving density profiles in the inset, which
is independent of the statistical parameter. At the beginning
anyons are confined in the central regime with an anyon per site.
After the harmonic trap is turned off, anyons expand in the
lattice and peaks appear at the boundary regime.

\section{conclusions}
In summary, in the present paper we have developed the exact
numerical method to deal with the dynamics of hard-core anyons
confined in the optical lattice superimposed with a weak harmonic
potential. By evaluating the exact time-dependent one-particle
Green's function, we obtain the density profiles, momentum
distributions, occupation distribution, and the occupation of the
lowest natural orbital in the full evolving period. It is shown
that the evolving property of density profiles is independent on
the statistical parameter of anyons.  As the harmonic trap is
turned off anyon shall expand in the optical lattice and
distribute gradually in the full lattice. The dynamical properties
of momentum distributions and occupation distributions of natural
orbitals depend on statistical parameter of anyons. In the Bose
limit, momentum distribution always displays the structure of
double peaks. As deviating from the Bose limit, the double peaks
become smaller and in the Fermi limit the double peaks disappear.
When the evolving time is long enough, anyons display the same
momentum distribution irrespective of statistical parameter in the
regime nearby zero momentum. If the anyon gas is initially in the
Mott regime, during the full evolving period  it shall be always
in the Mott regime but at the particular momentum position double
peaks appear. When the harmonic trap is replaced with a weaker
one, the density profiles of anyons shall exhibit the breathing
behavior. The momentum distributions of bosons alternatively
display the structure of single peak and the structure of double
peaks, while fermions  contract and expand in the $k$-space with
the time evolution. For anyon interpolating between these two
limit, momentum distribution also repeats its behavior at regular
intervals dependent on the statistical parameter. The interval
time is the longest for fermions and is the shortest for bosons.
The occupation distributions of natural orbitals do not change
qualitatively with the time evolution but the occupation of the
lowest natural orbital evolves with time for anyon gas deviating
the Fermi limit. When the harmonic trap is turned off it increases
and gradually arrives at the biggest value. When the harmonic trap
becomes weaker rather than being turned off, it shall fluctuate
with the time evolution.

\begin{acknowledgments}
This work was supported by NSF of China under Grants No. 11004007,
No.11174360 and No.10974234 and ``the Fundamental Research Funds
for the Central Universities." We thank A. del Campo for helpful
communications.
\end{acknowledgments}


\begin{references}
\bibitem{anyons}  J. M. Leinaasand and J. Myrheim, Nuovo Cimento {\bf 37B}, 1 (1977); F. Wilczek, Phys. Rev. Lett. {\bf 49}, 957 (1982).

\bibitem{Laughlin}  R. B. Laughlin, Phys. Rev. Lett. {\bf 50}, 1395 (1983).

\bibitem{Halperin}  B. I. Halperin, Phys. Rev. Lett. {\bf 52}, 1583 (1984).

\bibitem{Camino}  F. E. Camino, W. Zhou and V. J. Goldman, Phys. Rev. B {\bf 72}, 075342 (2005).

\bibitem{SuperConductor}  B. I. Halperin, Phys. Rev. Lett. {\bf 52}, 1583 (1984); F. Wilczek, Fractional Statistics and Anyon Superconductivity,
(World Scientific, Singapore 1990).

\bibitem{YSWu}  Y.-S. Wu and Y. Yu, Phys. Rev. Lett. {\bf 75}, 890 (1995).

\bibitem{ZNCHa}  Z. N. C. Ha, Phys. Rev. Lett. {\bf 73}, 1574 (1994); M. V. N. Murthy and R. Shankar, ibid. {\bf 73}, 3331 (1994); Z. N. C. Ha,
Nucl. Phys. B {\bf 435}, 604 (1995).

\bibitem{RMP2008}  C. Nayak, S. H. Simon, A. Stern, M. Freedman, and S. Das Sarma, Rev. Mod. Phys. {\bf 80}, 1083 (2008) .

\bibitem{Kitaev}  A. Y. Kitaev, Ann. of Phys. {\bf 303}, 2 (2003).

\bibitem{PZoller}  B. Paredes, P. Fedichev, J. I. Cirac, and P. Zoller, Phys. Rev. Lett. {\bf87}, 010402 (2001).
\bibitem{DuanLM} L.-M. Duan, E. Demler, and M. D. Lukin, Phys. Rev. Lett. {\bf91}, 090402 (2003);
A. Micheli, G. K. Brennen, and P. Zoller, Nature Phys. {\bf 2}, 341 (2006)��

\bibitem{OL} C.-W. Zhang, V. W. Scarola, Sumanta Tewari, and S. Das Sarma, Proc. Natl. Acad. Sci. USA {\bf 104}, 18415 (2007);
J.-K. Pachos, Ann. of Phys. {\bf 322}, 1254 (2007); M. Aguado, G. K.
Brennen, F. Verstraete, and J. I. Cirac, Phys. Rev. Lett. {\bf101},
260501 (2008).

\bibitem{NatureComm}  T. Keilmann, S. Lanzmich, I. McCulloch, and M. Roncaglia, Nature Communications, {\bf 2}, 361 (2011).

\bibitem{1D} M. A. Cazalilla, R. Citro, T. Giamarchi, E. Orignac, and M. Rigol, Rev. Mod. Phys \textbf{83}, 1405 (2011);
M. Olshanii, Phys. Rev. Lett. \textbf{81}, 938 (1998); D. S. Petrov,
G. V. Shlyapnikov, and J. T. M. Walraven, Phys. Rev. Lett.
\textbf{85}, 3745 (2000); V. Dunjko, V. Lorent, and M. Olshanii,
Phys. Rev. Lett. \textbf{86}, 5413 (2001).

\bibitem{Ketterler}  N. J. van Druten and W. Ketterle, Phys. Rev. Lett. {\bf 79}, 549 (1997).

\bibitem{Paredes}  B. Paredes, A. Widera, V. Murg, O. Mandel, S. F\"{o}lling, I. Cirac, G. V. Shlyapnikov, T. W. H\"{a}nsch, and I. Bloch,
Nature {\bf 429}, 277 (2004).

\bibitem{Toshiya}  T. Kinoshita, T. Wenger and D. S. Weiss, Science {\bf 305}, 1125 (2004).

\bibitem{STG} E. Haller, M. Gustavsson, M. J. Mark, J. G. Danzl, R. Hart, G. Pupillo, H.-C. N\"{a}gerl,Science \textbf{325}, 1224 (2009).

%\bibitem{Wilczek}  F. Wilczek, Phys. Rev. Lett. 49, 957 (1982).

\bibitem{Haldane}  F. D. M. Haldane, Phys. Rev. Lett. {\bf 67}, 937 (1991).

\bibitem{WangZD}  J. X. Zhu and Z. D. Wang, Phys. Rev. A {\bf 53}, 600 (1996).

\bibitem{Kundu99}  A. Kundu, Phys. Rev. Lett. {\bf 83}, 1275 (1999).

\bibitem{Girardeau06}  M. D. Girardeau, Phys. Rev. Lett. {\bf 97}, 210401
(2006).

\bibitem{Batchelor}  M. T. Batchelor, X. W. Guan, N. Oelkers, Phys. Rev. Lett. {\bf 96}, 210402 (2006);
M. T. Batchelor, X. W. Guan, J. S. He, J. Stat. Mech. P03007 (2007);
M. T. Batchelor, X. W. Guan, Phys. Rev. B {\bf 74}, 195121 (2006);
M. T. Batchelor, A. Foerster, X. W. Guan, J. Links, and H. Q. Zhou,
J. Phys. A: Math. Theor. {\bf 41}, 465201 (2008).

\bibitem{Patu07}  O. I. Patu, V. E. Korepin and D. V. Averin, J. Phys. A {\bf 40}, 14963 (2007).

%\bibitem{Batchelor07}  M. T. Batchelor, X. W. Guan, J. S. He, J. Stat. Mech. P03007 (2007).
%
%\bibitem{Batchelor06PRB}  M. T. Batchelor, X. W. Guan, Phys. Rev. B {\bf 74}, 195121 (2006).

\bibitem{Calabrese}  P. Calabrese and M. Mintchev, Phys. Rev. B {\bf 75}, 233104 (2007).

\bibitem{Cabra}  R. Santachiara, R. F. Stauffer and D. Cabra, J. Stat. Mech. L05003 (2007).

\bibitem{HaoPRA78}  Y. Hao, Y. Zhang, and S. Chen, Phys. Rev. A {\bf 78}, 023631 (2008).
\bibitem{HaoHCA79}  Y. Hao, Y. Zhang, and S. Chen, Phys. Rev. A {\bf 79}, 043633 (2009).

\bibitem{anyonTG}  R. Santachiara and P. Calabrese, J. Stat. Mech. P06005 (2008).

\bibitem{Patu08}  O. I. Patu, V. E. Korepin, and D. V. Averin, J. Phys. A
{\bf 41}, 145006 (2008); J. Phys. A: Math. Theor. {\bf 41} 255205 (2008).

\bibitem{Campo}  A. del Campo, Phys. Rev. A {\bf 78}, 045602 (2008).

\bibitem{Amico}  L. Amico, A. Osterloh, and U. Eckern, Phys. Rev. B {\bf 58}, R1703 (1998).

%\bibitem{Batchelor}  M. T. Batchelor, A. Foerster, X. W. Guan, J. Links, and H. Q. Zhou, J. Phys. A: Math. Theor. {\bf 41}, 465201 (2008).

%\bibitem{SWAP}  M. Anderlini, P. J. Lee, B. L. Brown, J. Sebby-Strabley, W.
%D. Phillips and J. V. Porto1, Nature {\bf 448}, 452 (2007).
%
%
%
%\bibitem{Aguado}  M. Aguado, G. K. Brennen, F. Verstraete, J. I. Cirac,
%Phys. Rev. Lett. {\bf 101}, 260501 (2008).
%
%\bibitem{Jiang}  L. Jiang, G. K. Brennen, A. V. Gorshkov, K. Hammerer, M.
%Hafezi, E. Demler, M. D. Lukin and P. Zoller, Nature Physics, {\bf 4}, 282
%(2008).


\bibitem{Girardeau}  M. D. Girardeau, J. Math. Phys. {\bf 6}, 516 (1960).

\bibitem{Hao06}  Y. Hao, Y. Zhang, J. Q. Liang, and S. Chen, Phys. Rev. A {\bf 73}, 063617 (2006).

\bibitem{Hao07}  Y. Hao, Y. Zhang, and S. Chen, Phys. Rev. A {\bf 76}, 063601 (2007).

\bibitem{Zoellner}  S. Z\"{o}llner, H.-D. Meyer, and P. Schmelcher, Phys. Rev. A {\bf 74}, 063611 (2006).

\bibitem{Deuretzbacher}  F. Deuretzbacher, K. Bongs, K. Sengstock, and D. Pfannkuche, Phys. Rev. A {\bf 75}, 013614 (2007); X. Yin, Y. Hao, S.
Chen, and Y. Zhang, Phys. Rev. A {\bf 78}, 013604 (2008).

\bibitem{Lenard}  A. Lenard, J. Math. Phys. {\bf 5}, 930 (1964); A. Lenard, J. Math. Phys. {\bf 7}, 1268 (1966).

\bibitem{Vaidya}  H. G. Vaidya and C. A. Tracy, Phys. Rev. Lett. {\bf 42}, 3 (1979) [Phys. Rev. Lett. {\bf 43}, E1540 (1979)]; H. G. Vaidya and
C. A. Tracy, J. Math. Phys. {\bf 20}, 2291 (1979).

\bibitem{MRigol}  M. Rigol and A. Muramatsu, Mod. Phys. Lett. B {\bf 19}, 861 (2005); M. Rigol and A. Muramatsu, Phys. Rev. A {\bf 70}, 031603 (2004);
M. Rigol and A. Muramatsu, Phys. Rev. A {\bf 72}, 013604 (2005).


\end{references}
\end{document}